\PassOptionsToPackage{unicode}{hyperref}
\PassOptionsToPackage{hyphens}{url}
\documentclass[
]{article}
\usepackage{xcolor}
\usepackage[margin=1in]{geometry}
\usepackage{amsmath,amssymb}
\usepackage{iftex}
\ifPDFTeX
  \usepackage[T1]{fontenc}
  \usepackage[utf8]{inputenc}
  \usepackage{textcomp} %
\else %
  \usepackage{unicode-math} %
  \defaultfontfeatures{Scale=MatchLowercase}
  \defaultfontfeatures[\rmfamily]{Ligatures=TeX,Scale=1}
\fi
\usepackage{lmodern}
\ifPDFTeX\else
\fi
\IfFileExists{upquote.sty}{\usepackage{upquote}}{}
\IfFileExists{microtype.sty}{%
  \usepackage[]{microtype}
  \UseMicrotypeSet[protrusion]{basicmath} %
}{}
\makeatletter
\@ifundefined{KOMAClassName}{%
  \IfFileExists{parskip.sty}{%
    \usepackage{parskip}
  }{%
    \setlength{\parindent}{0pt}
    \setlength{\parskip}{6pt plus 2pt minus 1pt}}
}{%
  \KOMAoptions{parskip=half}}
\makeatother
\usepackage{color}
\usepackage{fancyvrb}

\DefineVerbatimEnvironment{Highlighting}{Verbatim}{commandchars=\\\{\}}
\usepackage{framed}
\definecolor{shadecolor}{RGB}{248,248,248}
\newenvironment{Shaded}{\begin{snugshade}}{\end{snugshade}}

\newcommand{\AttributeTok}[1]{\textcolor[rgb]{0.13,0.29,0.53}{#1}}

\newcommand{\CommentTok}[1]{\textcolor[rgb]{0.56,0.35,0.01}{\textit{#1}}}

\newcommand{\ControlFlowTok}[1]{\textcolor[rgb]{0.13,0.29,0.53}{\textbf{#1}}}

\newcommand{\DecValTok}[1]{\textcolor[rgb]{0.00,0.00,0.81}{#1}}

\newcommand{\FloatTok}[1]{\textcolor[rgb]{0.00,0.00,0.81}{#1}}
\newcommand{\FunctionTok}[1]{\textcolor[rgb]{0.13,0.29,0.53}{\textbf{#1}}}

\newcommand{\NormalTok}[1]{#1}

\newcommand{\OtherTok}[1]{\textcolor[rgb]{0.56,0.35,0.01}{#1}}

\newcommand{\SpecialCharTok}[1]{\textcolor[rgb]{0.81,0.36,0.00}{\textbf{#1}}}

\newcommand{\StringTok}[1]{\textcolor[rgb]{0.31,0.60,0.02}{#1}}

\usepackage{longtable,booktabs,array}
\usepackage{calc} %
\usepackage{etoolbox}
\makeatletter
\patchcmd\longtable{\par}{\if@noskipsec\mbox{}\fi\par}{}{}
\makeatother
\IfFileExists{footnotehyper.sty}{\usepackage{footnotehyper}}{\usepackage{footnote}}
\makesavenoteenv{longtable}
\usepackage{graphicx}
\makeatletter
\newsavebox\pandoc@box
\newcommand*\pandocbounded[1]{%
  \sbox\pandoc@box{#1}%
  \Gscale@div\@tempa{\textheight}{\dimexpr\ht\pandoc@box+\dp\pandoc@box\relax}%
  \Gscale@div\@tempb{\linewidth}{\wd\pandoc@box}%
  \ifdim\@tempb\p@<\@tempa\p@\let\@tempa\@tempb\fi%
  \ifdim\@tempa\p@<\p@\scalebox{\@tempa}{\usebox\pandoc@box}%
  \else\usebox{\pandoc@box}%
  \fi%
}
\def\fps@figure{htbp}
\makeatother
\NewDocumentCommand\citeproctext{}{}
\NewDocumentCommand\citeproc{mm}{%
  \begingroup\def\citeproctext{#2}\cite{#1}\endgroup}
\makeatletter
 \let\@cite@ofmt\@firstofone
 \def\@biblabel#1{}
 \def\@cite#1#2{{#1\if@tempswa , #2\fi}}
\makeatother
\newlength{\cslhangindent}
\newlength{\csllabelwidth}
\newenvironment{CSLReferences}[2] %
 {\begin{list}{}{%
  \setlength{\itemindent}{0pt}
  \setlength{\leftmargin}{0pt}
  \setlength{\parsep}{0pt}
  \ifodd #1
   \setlength{\leftmargin}{\cslhangindent}
   \setlength{\itemindent}{-1\cslhangindent}
  \fi
  \setlength{\itemsep}{#2\baselineskip}}}
 {\end{list}}
\usepackage{calc}

\providecommand{\tightlist}{%
  \setlength{\itemsep}{0pt}\setlength{\parskip}{0pt}}
\usepackage{amsmath,amssymb}
\usepackage{booktabs}
\usepackage{hyperref}
\usepackage[margin=1in]{geometry}
\usepackage{fvextra}
\DefineVerbatimEnvironment{Highlighting}{Verbatim}{breaklines,commandchars=\\\{\}}
\makeatletter
\AtBeginDocument{%
  \renewcommand{\@author}{Keaven M. Anderson and Alison Pedley\\[0.3em]%
    \normalsize\textit{Merck \& Co., Inc., Rahway, NJ, USA}}%
}
\makeatother
\usepackage{bookmark}
\IfFileExists{xurl.sty}{\usepackage{xurl}}{} %
\hypersetup{
  pdftitle={Two-Stage Design with Sample Size Re-estimation using gsDesign},
  pdfauthor={Keaven M. Anderson; Alison Pedley},
  hidelinks,
  pdfcreator={LaTeX via pandoc}}

\title{Two-Stage Design with Sample Size Re-estimation using gsDesign}
\author{Keaven M. Anderson \and Alison Pedley}
\date{August 04, 2026}

\begin{document}
\maketitle
\begin{abstract}
The effect size and nuisance parameters needed to appropriately size a clinical trial are generally not adequately understood before a trial begins.
Through pre-planned early stopping rules, a conservatively planned group sequential design can effectively adapt the sample size for a clinical trial for a new treatment that is ineffective, very effective or minimally effective.
One potential issue with such designs is that a substantial number of patients can be enrolled after a data cutoff for an interim analysis while data are being entered, cleaned, analyzed and discussed.
A strategy of re-estimating the sample size at an interim analysis based on conditional power has been proposed to reduce somewhat this enrollment overrun issue. A purported advantage sometimes claimed is a smaller up-front planned sample size than a conservatively planned group sequential design.
We demonstrate derivation of 2-stage group sequential and conditional power designs using the gsDesign R package and suggest comparing designs with comparable power using expected sample size calculations.
While there are cases where conditional power designs may have small advantages, it is quite easy to derive very inefficient conditional power designs. This, along with the fact that a conditional power design may reveal something about the interim treatment effect, will often leave a group sequential design as the design of choice.
\end{abstract}

Note that gsDesign version \(\geq\) 3.10.0 is required to reproduce this paper.

\section{Introduction}\label{sec:intro}

The objective of this paper is to demonstrate a tool to derive and evaluate 2-stage designs that adapt sample size at an interim analysis based on conditional power. The rationale for conditional power designs is to `rescue' trials that appear to have less than the desired power based on an interim analysis. Such designs have been proposed for at least 25 years (\citeproc{ref-Bauer}{Bauer 1989}; \citeproc{ref-CuiHungWang}{Cui, Hung, and Wang 1999}; \citeproc{ref-ProschanHunsberger}{Proschan and Hunsberger 1995}) and refinements have been made to improve design properties while either maintaining the use of conditional power, e.g., (\citeproc{ref-LiuChi}{Liu and Chi 2001}; \citeproc{ref-GaoWareMehta}{Gao, Ware, and Mehta 2008}; \citeproc{ref-MehtaPocock}{Mehta and Pocock 2011}), or not (\citeproc{ref-Posch}{Posch, Bauer, and Brannath 2003}; \citeproc{ref-Lokhnygina}{Lokhnygina and Tsiatis 2008}; \citeproc{ref-Schmitz}{Schmitz 1993}). Since it is easy to produce an inefficient conditional power design (\citeproc{ref-JTStatMed}{Jennison and Turnbull 2003}), it seems valuable to consider the overall power gains and cost of said gains in terms of expected sample size. Using the gsDesign R package we compare power and sample size between a 2-stage conditional power design and group sequential designs. While we try to provide clarity, we also assume the reader will install the package and examine its extensive help files where some of the code can be further clarified.

While the focus here is on comparing designs with sample size re-estimation using conditional power versus group sequential designs, we also provide some overview here of the more general gsDesign package and related R packages.
The gsDesign package was first released in 2006 to derive group sequential designs.
It was substantially updated and rewritten through 2009 and has received additional functionality since that time, particularly in the area of trial designs for time-to-event endpoints.
Functionality for printing, \texttt{xtable} (for LaTeX support) and plotting emphasizing capabilities from the ggplot2 package are included.
A web interface is provided at \url{https://gsdesign.shinyapps.io/prod/}, including the capability to write code that can be saved and re-used with the gsDesign package.

The history of conditional power designs dating from 1989 was documented by Bauer et al. (\citeproc{ref-bauer2016twenty}{2016}).
Similar methods for two-stage adaptive designs are also implemented in the \texttt{adaptTest} package (\citeproc{ref-adaptTest}{Vandemeulebroecke 2009}) and estimation following adaptive designs is available in the \texttt{AGSDest} package (\citeproc{ref-HackBrannath}{Hack, Brannath, and Brueckner 2013}).
Wassmer and Brannath (\citeproc{ref-WassmerBrannath}{2016}) provide an extensive textbook treatment of group sequential and confirmatory adaptive designs.
Regulatory guidance on the topic has been provided by the U.S. FDA (\citeproc{ref-FDAADFinal}{Center for Drug Evaluation and Research and Center for Biologics Evaluation and Research 2019}) and EMA (\citeproc{ref-CHMPAD}{CHMP Efficacy Working Party 2007}).

\section{Example trial}\label{sec:example}

We will begin with a fixed design sample size for all of the methods considered.
We will assume 80\% power (Type II error 20\%) and 2.5\% one-sided Type I error for all cases.
Clinical trials with a control and experimental treatment arm will be considered.
For a continuous, normally distributed outcome and reasonably large sample sizes, the function \texttt{nNormal()} that assumes known variance should be adequate. We consider an example from Wang, Hung, and O'Neill (\citeproc{ref-WangHungOSIM2012}{2012}).
They compared normally distributed means with a standard deviation of 1 for treatments.
Early studies indicated a treatment benefit with a new, experimental therapy to be between 0.27 and 0.33.
We begin by computing the total required sample size for a new trial using the more optimistic of these assumptions with \(\delta = \mu_1 - \mu_2 = 0.33\).

\begin{Shaded}
\begin{Highlighting}[]
\CommentTok{\# Compute sample size with 1:1 randomization}
\NormalTok{nNor }\OtherTok{\textless{}{-}} \FunctionTok{nNormal}\NormalTok{(}\AttributeTok{delta1 =} \FloatTok{0.33}\NormalTok{, }\AttributeTok{sd =} \DecValTok{1}\NormalTok{, }\AttributeTok{alpha =} \FloatTok{0.025}\NormalTok{, }\AttributeTok{beta =} \FloatTok{0.2}\NormalTok{)}
\CommentTok{\# Round up to an even number}
\DecValTok{2} \SpecialCharTok{*} \FunctionTok{ceiling}\NormalTok{(nNor }\SpecialCharTok{/} \DecValTok{2}\NormalTok{)}
\end{Highlighting}
\end{Shaded}

\begin{verbatim}
## [1] 290
\end{verbatim}

For smaller sample sizes a routine based on the \(t\)-distribution from base R may be preferred, yielding a total sample size of
292 in this case: \texttt{power.t.test(delta\ =\ 0.33,\ power\ =\ 0.80)}.
Note that \texttt{nNormal()} allows unequal randomization and designing for non-inferiority, while these capabilities are not built into \texttt{power.t.test()}.

If, in truth, the effect size were the smaller \(\delta = 0.27\) instead of \(\delta = 0.33\), the sample size required to power the trial would be 432 instead of 290. If the effect size were correctly specified as \(\delta = 0.33\), but standard deviation were \(\sigma = 1.5\) instead of \(\sigma = 1\), the required sample size would be 650. Methods in subsequent sections will attempt to adapt from the original assumptions at an interim analysis to power the trial appropriately.

The efficient estimate for the parameter of interest in this case is \(\hat{\delta} = \bar{X}_1 - \bar{X}_2\) where \(\bar{X}_j\) is the sample mean after \(n/2\) observations for group \(j\), \(j = 1, 2\).
For testing in this case, we would use \(Z = \sqrt{n}(\bar{X}_1 - \bar{X}_2) / (2\sigma) = \sqrt{n}\hat{\delta} / (2\sigma)\) which has a standard normal distribution with mean \(\sqrt{n}\hat{\delta} / (2\sigma)\).
The standardized effect size \(\theta\) is defined as \(\delta / (2\sigma)\) for a two-sample normal test which implies \(\hat{\theta} = \hat{\delta} / (2\sigma)\) and
\[Z \sim \text{Normal}(\sqrt{n}\theta, 1).\]
The sample size to achieve power \(1 - \beta\) with one-sided Type I error \(\alpha\) is
\[\left(\frac{\Phi^{-1}(1-\alpha) + \Phi^{-1}(1-\beta)}{\theta}\right)^2\]
where \(\Phi^{-1}(\cdot)\) is the inverse of the cumulative standard normal distribution function.

\section{Group sequential design}\label{sec:gsd}

\begin{figure}

{\centering \includegraphics{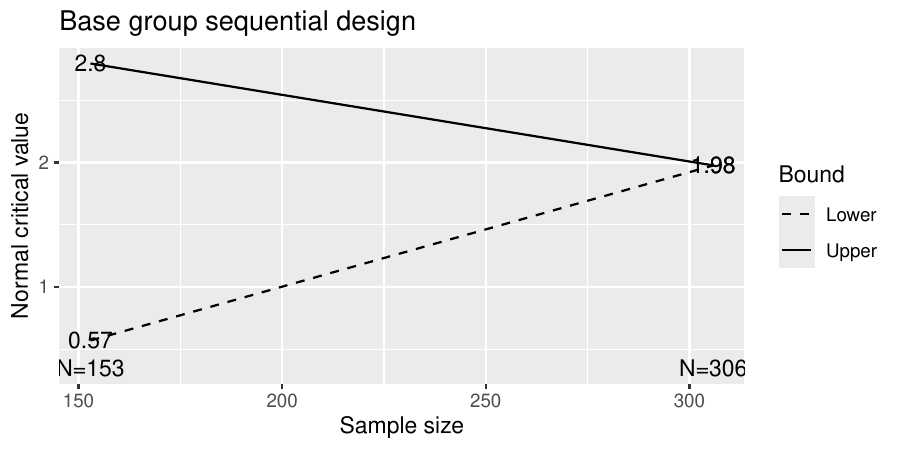} 

}

\caption{Group sequential design sample size and normal test statistic bounds at interim and final analysis for a standard deviation of 1 and a difference in group means of 0.33.}\label{fig:gsD}
\end{figure}

Group sequential design will adapt to different effect sizes if a minimum effect size of interest can be specified and early stopping for efficacy or futility is used to adapt sample size. With a 2-stage group sequential design, we perform a single interim analysis and stop the trial for positive efficacy if the interim test statistic is very positive, stop for futility if the results are discouraging, and otherwise continue the trial to the end.
The basic algorithm (\citeproc{ref-JTBook}{Jennison and Turnbull 2000}) inflates a sample size from a fixed design to produce a group sequential design as derived with the code below and shown in Figure \ref{fig:gsD}.
Given the cost of powering for the more conservative effect size of 0.27, assume we wish to start with the more optimistic treatment benefit of 0.33.
The following code plots the expected sample size by underlying treatment effect assuming an enrollment overrun of 75 enrolled beyond those with outcome data available at the interim analysis.
The values \texttt{k\ =\ 2} and \texttt{timing\ =\ 0.5}, respectively, specify two analyses with the interim analysis including half of the planned observations.
The design bounds and sample size are derived using spending functions (\citeproc{ref-LanDeMets}{Lan and DeMets 1983}).
Specifically, we use the power spending function of Kim and DeMets (\citeproc{ref-KimDeMets}{1987}) (\(\alpha(t) = \alpha t^\rho\)).
The upper (efficacy) spending function is specified in the argument \texttt{sfu} and the lower (futility) spending function in \texttt{sfl}; the same spending function family is used for both in order to simply make bounds for alternate designs comparable. The value of the parameter \texttt{sfupar} here was chosen to match the commonly used O'Brien and Fleming (\citeproc{ref-OF}{1979}) bound, while the lower bound (parameter \texttt{sflpar}) is intended to provide a reasonable probability of stopping the trial for futility at the interim analysis if the underlying treatment effect is less than targeted.
Assuming the default one-sided Type I error \texttt{alpha\ =\ 0.025} and \(\rho = 3.275\) in the power spending function chosen, the nominal Type I error at the interim analysis half-way through the trial (\(t = 0.5\)) is \(0.025 \times 0.5^{3.275} =\) 0.0026 which can be seen if you enter the command \texttt{gsBoundSummary(gsD)}.
You can also see that the probability of crossing the futility bound is \(0.2 \times 0.5^{1.5} =\) 0.0707, where \(\beta = 0.2\) is the Type II error, \(\rho = 1.5\) is the power spending function parameter and, again, \(t = 0.5\) to indicate the analysis is done after half of the final planned sample size is available.

\begin{Shaded}
\begin{Highlighting}[]
\CommentTok{\# Extend the fixed design sample size to a group sequential design}
\NormalTok{gsD }\OtherTok{\textless{}{-}} \FunctionTok{gsDesign}\NormalTok{(}
  \AttributeTok{k =} \DecValTok{2}\NormalTok{, }\AttributeTok{timing =} \FloatTok{0.5}\NormalTok{, }\AttributeTok{beta =} \FloatTok{0.2}\NormalTok{, }\AttributeTok{alpha =} \FloatTok{0.025}\NormalTok{,}
  \AttributeTok{n.fix =}\NormalTok{ nNor, }\AttributeTok{delta1 =} \FloatTok{0.33}\NormalTok{, }\AttributeTok{overrun =} \DecValTok{75}\NormalTok{,}
  \AttributeTok{sfu =}\NormalTok{ sfPower, }\AttributeTok{sfl =}\NormalTok{ sfPower, }\AttributeTok{sfupar =} \FloatTok{3.275}\NormalTok{, }\AttributeTok{sflpar =} \FloatTok{1.5}
\NormalTok{)}
\CommentTok{\# Plot z{-}value bounds for the group sequential design}
\FunctionTok{plot}\NormalTok{(gsD, }\AttributeTok{cex =} \FloatTok{0.8}\NormalTok{, }\AttributeTok{main =} \StringTok{"Base group sequential design"}\NormalTok{)}
\end{Highlighting}
\end{Shaded}

\begin{figure}

{\centering \includegraphics{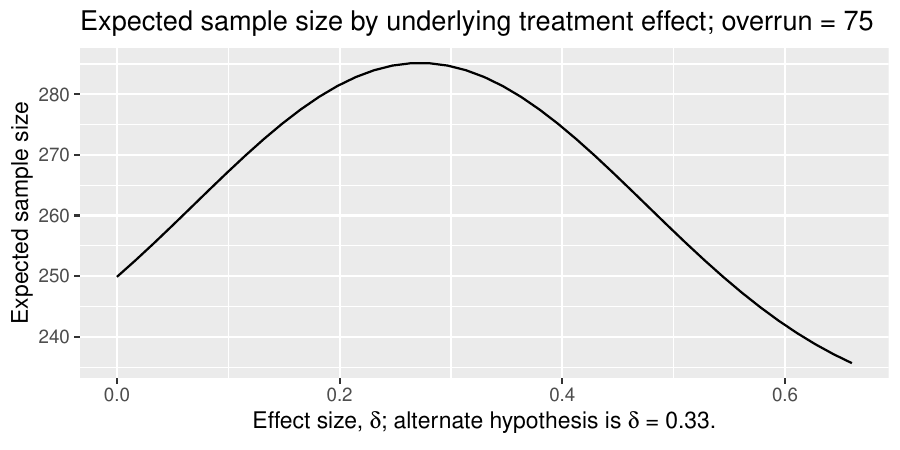} 

}

\caption{Expected sample size by treatment effect for the planned 2-stage group sequential design powered for a treatment effect of $\delta = 0.33$ assuming a standard deviation of 1 and enrollment overrun of 75 at the interim analysis.}\label{fig:gsDen}
\end{figure}

If the interim can be conducted promptly after the interim patient set is enrolled, this can result in substantial savings compared to the fixed design sample size of 290; Figure \ref{fig:gsDen} shows the expected sample size with the assumed interim analysis overrun of 75 patients.

\begin{Shaded}
\begin{Highlighting}[]
\CommentTok{\# Plot expected sample size by underlying treatment effect}
\FunctionTok{plot}\NormalTok{(gsD, }\AttributeTok{plottype =} \StringTok{"asn"}\NormalTok{,}
     \AttributeTok{main =} \StringTok{"Expected sample size by underlying treatment effect; overrun = 75"}\NormalTok{,}
     \AttributeTok{xlab =} \FunctionTok{expression}\NormalTok{(}\FunctionTok{paste}\NormalTok{(}\StringTok{"Effect size, "}\NormalTok{, delta,}
                             \StringTok{"; alternate hypothesis is "}\NormalTok{, delta, }\StringTok{" = 0.33."}\NormalTok{)))}
\end{Highlighting}
\end{Shaded}

We also show how to print the expected sample sizes given \(\delta = 0\), \(0.27\) and \(0.33\), respectively.

\begin{Shaded}
\begin{Highlighting}[]
\CommentTok{\# Print power for delta = 0, .27, .33 recalling that standardized effect size}
\CommentTok{\# theta = delta / 2; the gsProbability function can add values of theta into a}
\CommentTok{\# gsDesign object}
\FunctionTok{gsProbability}\NormalTok{(}\AttributeTok{d =}\NormalTok{ gsD, }\AttributeTok{theta =} \FunctionTok{c}\NormalTok{(}\DecValTok{0}\NormalTok{, }\FloatTok{0.27}\NormalTok{, }\FloatTok{0.33}\NormalTok{) }\SpecialCharTok{/} \DecValTok{2}\NormalTok{)}\SpecialCharTok{$}\NormalTok{en}
\end{Highlighting}
\end{Shaded}

\begin{verbatim}
## [1] 249.8941 285.1678 282.8383
\end{verbatim}

\begin{figure}

{\centering \includegraphics{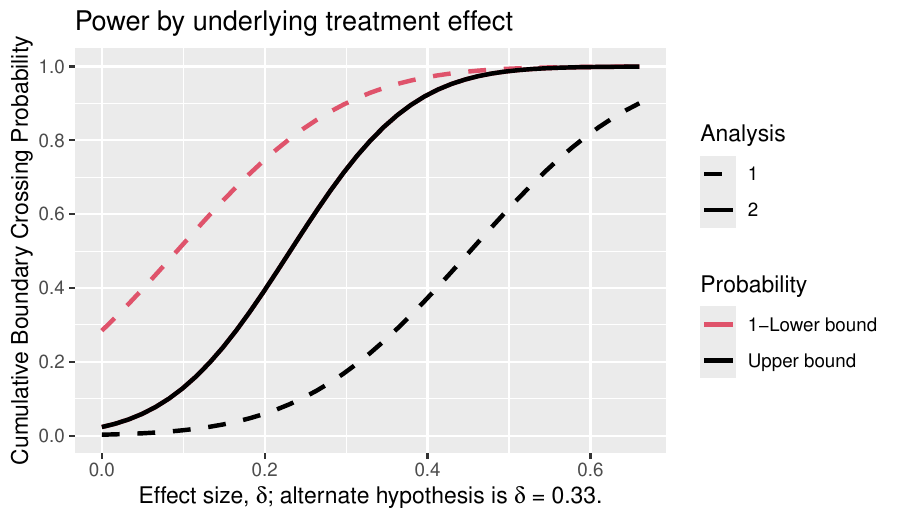} 

}

\caption{Power by treatment effect for the planned 2-stage group sequential design powered for a treatment effect of $\delta = 0.33$ assuming a standard deviation of 1.}\label{fig:gspower}
\end{figure}

We can also plot power as shown in Figure \ref{fig:gspower}. The solid black line shows that there is a sharp drop in overall study power (positive result at interim or final analysis) from 80\% with \(\delta = 0.33\) to 63\% if the less optimistic \(\delta = 0.27\) holds. The dashed black line shows the probability of crossing the efficacy bound at the interim analysis, while the dashed red line shows one minus the probability of crossing the futility bound at the interim analysis.
Thus, the lines divide the probability of all possible boundary crossing possibilities for each underlying effect size.

\begin{Shaded}
\begin{Highlighting}[]
\CommentTok{\# Plot power by underlying treatment effect}
\FunctionTok{plot}\NormalTok{(gsD, }\AttributeTok{plottype =} \StringTok{"power"}\NormalTok{,}
     \AttributeTok{main =} \StringTok{"Power by underlying treatment effect"}\NormalTok{,}
     \AttributeTok{xlab =} \FunctionTok{expression}\NormalTok{(}\FunctionTok{paste}\NormalTok{(}\StringTok{"Effect size, "}\NormalTok{, delta,}
                             \StringTok{"; alternate hypothesis is "}\NormalTok{, delta, }\StringTok{" = 0.33."}\NormalTok{)))}
\end{Highlighting}
\end{Shaded}

A group sequential design fully powered for \(\delta = 0.27\) would be one way of ensuring adequate power for this smaller effect size.
Another way of ensuring adequate power for the smaller effect size would be to design a group sequential trial with an interim analysis after 50\% of the trial has been analyzed and adapt the sample size up if interim results are in the continuation region of the group sequential design and ``sufficiently encouraging.'' This is done with the thought of improving the power of the design without incurring the additional associated expense if such an increase is not needed. The discussion that follows shows some suggested adaptation methods from the literature as well as a method to evaluate the impact of such strategies on the expected sample size (cost) required and study power.

\section{Two-stage sample size re-estimation}\label{sec:ssr}

The notation for the above group sequential design is \(n_1\) observations at the interim analysis, an additional \(n_2\) for a total of \(N_2 = n_1 + n_2\) at the final analysis, and for analyses \(i = 1, 2\), futility bounds \(a_i\) and efficacy bounds \(b_i\).
We assume test statistics \(Z_1\) and \(Z_2\) that approximately follow a bivariate normal distribution with means \(\sqrt{n_1}\theta\) and \(\sqrt{N_2}\theta\), respectively, variances of 1, \(i = 1, 2\), and covariance \(\sqrt{n_1/N_2}\).
Let \(w_i = \sqrt{n_i/N_2}\) for \(i = 1, 2\) and assume \(Y_2 \sim \text{Normal}(\sqrt{n_2}\theta, 1)\).
Letting
\[Z_2 = w_1 Z_1 + w_2 Y_2\]
produces the same multivariate normal distribution for \(Z_1\) and \(Z_2\).
For our specific example examining the difference in means of two normal samples with a common variance and sample sizes, we have \(Z_1\) based on the first \(n_1/2\) observations for each group and \(Y_2\) based on the next \(n_2/2\) observations for each group.

We consider a two-stage adaptive design here that is a generalization building on the group sequential design just outlined.
The approach taken assumes a second stage sample size \(n_2(Z_1)\), and a conditionally independent normally distributed random variable \[Y_2(Z_1) | (Z_1 = c) \sim \text{Normal}\left(\sqrt{n_2(c)}\theta, 1\right).\]
We denote the total sample size as
\[N_2(Z_1) = n_1 + n_2(Z_1).\]
Note that the functions \(N_2(\cdot)\), \(n_2(\cdot)\) and the random variable \(Y_2(Z_1)\) are all generalizations from their respective simplest forms \(N_2\), \(n_2\) and \(Y_2\) from the group sequential design.
We also assume that the trial is declared positive if \(Y_2(Z_1) \geq c(Z_1, n_2(Z_1))\) for functions \(c(\cdot, \cdot)\) and \(n_2(\cdot)\). For the group sequential case, since the trial is declared positive if \(Z_2 \geq b_2\), we have
\begin{equation}
c(Z_1, n_2(Z_1)) = \frac{b_2 - w_1 Z_1}{w_2}. \label{eq:gsw}
\end{equation}
We assume the trial stops at stage 1 for futility with sample size \(n_1\) included in the analysis if \(Z_1 < a_1\) or for efficacy if \(Z_1 \geq b_1\). Letting \(\phi(\cdot)\) denote the standard normal density function, the probability of a positive trial is
\begin{align}
\alpha(\theta) &= P_\theta\{Z_1 \geq b_1\} + P_\theta\{Y_2(Z_1) \geq c(Z_1, n_2(Z_1))\} \nonumber\\
&= 1 - \Phi(b_1 - \sqrt{n_1}\theta) + \int_{a_1}^{b_1} \phi(z_1 - \sqrt{n_1}\theta)\left(1 - \Phi\left(c(z_1, n_2(z_1)) - \sqrt{n_2(z_1)}\theta\right)\right) dz_1. \label{eq:cprob}
\end{align}
Equation \eqref{eq:cprob} is general in that it can apply to a 2-stage group sequential design, various conditional power designs (\citeproc{ref-CuiHungWang}{Cui, Hung, and Wang 1999}; \citeproc{ref-ProschanHunsberger}{Proschan and Hunsberger 1995}; \citeproc{ref-LehmacherWassmer}{Lehmacher and Wassmer 1999}; \citeproc{ref-LiuChi}{Liu and Chi 2001}; \citeproc{ref-ChenDeMetsLan}{Chen, DeMets, and Lan 2004}; \citeproc{ref-MehtaPocock}{Mehta and Pocock 2011}), or other sample size adaptation methods such as Posch, Bauer, and Brannath (\citeproc{ref-Posch}{2003}).
The optimal two-stage design of Lokhnygina and Tsiatis (\citeproc{ref-Lokhnygina}{2008}) also takes this form, although computing \(n_2(z_1)\) requires dynamic programming.

Next we focus on the choice of \(c(z_1, n_2(z_1))\) where \(z_1\) is assumed to be an observed value of \(Z_1\). There are two options we will consider here, both of which use a weighted combination of \(Z_1\) and \(Y_2(Z_1)\) taking a generalization of the form in equation \eqref{eq:gsw}:
\begin{equation}
c(Z_1, n_2(Z_1)) = \frac{b_2 - Z_1 w_1(Z_1)}{w_2(Z_1)}. \label{eq:czn}
\end{equation}

\begin{itemize}
\item
  Using a combination test weighting \(Z_1\) and \(Y_2(Z_1)\) as planned in the underlying group sequential design (\citeproc{ref-CuiHungWang}{Cui, Hung, and Wang 1999}) guarantees control of Type I error by down-weighting observations composing \(Y_2(Z_1)\) if the sample size is increased. Noting that we are using the group sequential design \(n_2\), not the more general \(n_2(Z_1)\), the weights for \(i = 1, 2\) are:
  \begin{equation}
  w_i(z_i) = \sqrt{n_i / (n_1 + n_2)}. \label{eq:wfixed}
  \end{equation}
  Defining \(Z_2(Z_1) = w_1 Z_1 + w_2 Y_2(Z_1)\) we note that the pair \((Z_1, Z_2(Z_1))\) has the same distribution as the pair \((Z_1, Z_2)\) from the group sequential design and, thus, the testing has not changed from the original group sequential test.
\item
  Using a combination that corresponds to testing at the second analysis with a sufficient statistic, equally weighting observations before and after the interim analysis. This requires some restrictions in order to ensure Type I error is controlled (\citeproc{ref-ChenDeMetsLan}{Chen, DeMets, and Lan 2004}; \citeproc{ref-GaoWareMehta}{Gao, Ware, and Mehta 2008}). Here we have
  \begin{align}
  w_1(z_1) &= \sqrt{n_1 / (n_1 + n_2(z_1))} \label{eq:w1z1}\\
  w_2(z_1) &= \sqrt{n_2(z_1) / (n_1 + n_2(z_1))}. \label{eq:w2z1}
  \end{align}
\end{itemize}

Now we need to specify \(n_2(z_1)\). For the methods presented here, this will be based on conditional power. Let \(\beta^\star\) represent a target conditional Type II error for the trial conditioning on the interim test statistic \(Z_1 = z_1\); normally \(\beta^\star\) equals the Type II error \(\beta\) planned for the underlying group sequential design. The conditional power is computed under the assumption that \(\theta = \theta(z_1)\). Some authors propose an efficient estimator such as \(\theta(z_1) = \hat{\theta}_1 \approx z_1 / \sqrt{n_1}\) (\citeproc{ref-CuiHungWang}{Cui, Hung, and Wang 1999}; \citeproc{ref-ProschanHunsberger}{Proschan and Hunsberger 1995}; \citeproc{ref-ChenDeMetsLan}{Chen, DeMets, and Lan 2004}; \citeproc{ref-GaoWareMehta}{Gao, Ware, and Mehta 2008}; \citeproc{ref-MehtaPocock}{Mehta and Pocock 2011}). However, Liu and Chi (\citeproc{ref-LiuChi}{2001}) suggest using the originally specified value for the alternate hypothesis for which the group sequential trial was powered, \(\theta(z_1) = \theta_1\). In any case, letting \(\Phi(\cdot)\) represent the standard normal cumulative distribution function, we wish to select \(n_2(z_1)\) to satisfy
\begin{align}
\beta^\star &= P\{Y_2(z_1) < c(z_1, n_2(z_1)) | \theta = \theta(z_1)\} \nonumber\\
&= \Phi\left(c(z_1, n_2(z_1)) - \sqrt{n_2(z_1)}\theta(z_1)\right). \label{eq:betastar}
\end{align}
To satisfy this, from equations \eqref{eq:czn} and \eqref{eq:betastar} we must have
\begin{equation}
n_2(z_1) = \left(\frac{c(z_1, n_2(z_1)) + \Phi^{-1}(1 - \beta^\star)}{\theta(z_1)}\right)^2. \label{eq:cpn}
\end{equation}
Note that in the case of equation \eqref{eq:wfixed} that equation \eqref{eq:cpn} can be used to directly compute \(n_2(\cdot)\) to achieve the desired conditional power since \(c(z_1, n_2(z_1))\) does not depend on the function \(n_2(\cdot)\).
When \(c(z_1, n_2(z_1))\) depends on \(n_2(z_1)\) in equations \eqref{eq:w1z1}--\eqref{eq:w2z1}, fixed point iteration can be simply applied as follows:

\begin{enumerate}
\def\labelenumi{\arabic{enumi}.}
\tightlist
\item
  Initialize \(n_2(z_1)\) using equations \eqref{eq:czn} and \eqref{eq:wfixed}.
\item
  Compute \(c(z_1, n_2(z_1))\) based on the computed \(n_2\)-values and desired \(c(\cdot)\)-function.
\item
  Compute \(n_2(z_1)\) based on equation \eqref{eq:cpn}.
\item
  Repeat steps 2 and 3 until \(n_2\) converges to a solution.
\end{enumerate}

This normally can be completed adequately in a small, fixed number of iterations.

\section{Code for conditional power at a group sequential design interim analysis}\label{sec:condpower}

\begin{figure}

{\centering \includegraphics{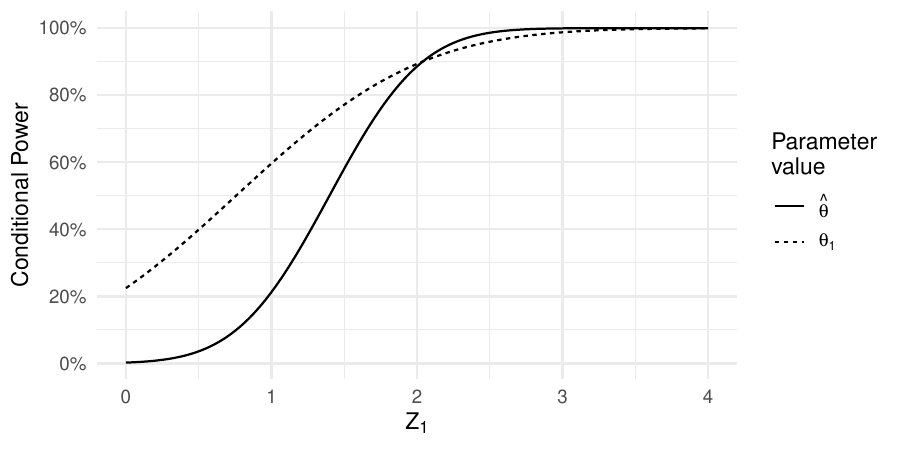} 

}

\caption{Conditional power based on interim test statistic and assumed parameter value.}\label{fig:condPower}
\end{figure}

Before deriving adaptive designs based on conditional power, we show the conditional power for the group sequential design under consideration (without sample size adaptation) assuming the trial continues past the interim analysis.
This is computed using the function \texttt{condPower()}. The default assumes the true treatment effect for the second stage of the trial is the estimated treatment effect based on data from the first part of the trial, \(\hat{\theta}_1 \approx z_1 / \sqrt{n_1}\). The following code generates Figure \ref{fig:condPower} which demonstrates the conditional power for the group sequential design assuming either 1) the interim treatment effect \(\hat{\theta}_1\) or 2) the treatment effect \(\theta_1\) under the alternate hypothesis. Note that the conditional power based on the observed effect size equals the design power of 80\% when \(z_1 \approx 1.8\) and drops off sharply for smaller values of \(z_1\).

\begin{Shaded}
\begin{Highlighting}[]
\CommentTok{\# Select z{-}statistic values at interim to evaluate}
\NormalTok{z }\OtherTok{\textless{}{-}} \FunctionTok{seq}\NormalTok{(}\DecValTok{0}\NormalTok{, }\DecValTok{4}\NormalTok{, }\FloatTok{0.0125}\NormalTok{)}
\CommentTok{\# 2nd stage sample size is difference between total at final and interim}
\NormalTok{n2 }\OtherTok{\textless{}{-}}\NormalTok{ gsD}\SpecialCharTok{$}\NormalTok{n.I[}\DecValTok{2}\NormalTok{] }\SpecialCharTok{{-}}\NormalTok{ gsD}\SpecialCharTok{$}\NormalTok{n.I[}\DecValTok{1}\NormalTok{]}
\CommentTok{\# Compute conditional power based on observed treatment effect}
\NormalTok{cphat }\OtherTok{\textless{}{-}} \FunctionTok{condPower}\NormalTok{(}\AttributeTok{z1 =}\NormalTok{ z, }\AttributeTok{n2 =}\NormalTok{ n2, }\AttributeTok{x =}\NormalTok{ gsD)}
\CommentTok{\# Compute conditional power based on H1 treatment effect}
\NormalTok{cp1 }\OtherTok{\textless{}{-}} \FunctionTok{condPower}\NormalTok{(}\AttributeTok{z1 =}\NormalTok{ z, }\AttributeTok{n2 =}\NormalTok{ n2, }\AttributeTok{x =}\NormalTok{ gsD, }\AttributeTok{theta =}\NormalTok{ gsD}\SpecialCharTok{$}\NormalTok{delta)}

\CommentTok{\# Combine into a data frame and plot}
\NormalTok{cp }\OtherTok{\textless{}{-}} \FunctionTok{bind\_rows}\NormalTok{(}
  \FunctionTok{tibble}\NormalTok{(}\AttributeTok{z1 =}\NormalTok{ z, }\AttributeTok{CP =}\NormalTok{ cphat, }\AttributeTok{Parameter =} \StringTok{"Observed"}\NormalTok{),}
  \FunctionTok{tibble}\NormalTok{(}\AttributeTok{z1 =}\NormalTok{ z, }\AttributeTok{CP =}\NormalTok{ cp1, }\AttributeTok{Parameter =} \StringTok{"Planned"}\NormalTok{)}
\NormalTok{)}
\FunctionTok{ggplot}\NormalTok{(cp, }\FunctionTok{aes}\NormalTok{(}\AttributeTok{x =}\NormalTok{ z1, }\AttributeTok{y =}\NormalTok{ CP, }\AttributeTok{lty =}\NormalTok{ Parameter)) }\SpecialCharTok{+}
  \FunctionTok{geom\_line}\NormalTok{() }\SpecialCharTok{+}
  \FunctionTok{ylab}\NormalTok{(}\StringTok{"Conditional Power"}\NormalTok{) }\SpecialCharTok{+}
  \FunctionTok{xlab}\NormalTok{(}\FunctionTok{expression}\NormalTok{(Z[}\DecValTok{1}\NormalTok{])) }\SpecialCharTok{+}
  \FunctionTok{scale\_y\_continuous}\NormalTok{(}\AttributeTok{labels =}\NormalTok{ percent, }\AttributeTok{breaks =}\NormalTok{ (}\DecValTok{0}\SpecialCharTok{:}\DecValTok{5}\NormalTok{) }\SpecialCharTok{/} \DecValTok{5}\NormalTok{) }\SpecialCharTok{+}
  \FunctionTok{scale\_linetype\_discrete}\NormalTok{(}
    \AttributeTok{name =} \StringTok{"Parameter}\SpecialCharTok{\textbackslash{}n}\StringTok{value"}\NormalTok{,}
    \AttributeTok{labels =} \FunctionTok{c}\NormalTok{(}\FunctionTok{expression}\NormalTok{(}\FunctionTok{hat}\NormalTok{(theta)), }\FunctionTok{expression}\NormalTok{(theta[}\DecValTok{1}\NormalTok{]))}
\NormalTok{  ) }\SpecialCharTok{+}
  \FunctionTok{theme\_minimal}\NormalTok{()}
\end{Highlighting}
\end{Shaded}

While the group sequential design will perform as designed, the presumed ``problem'' demonstrated above is that the conditional power for the trial given the interim treatment effect and test statistic is often lower than the originally planned power.

\subsection{Scale equivalences at an interim analysis}\label{sec:scales}

Since \(Z_1\) is the primary output of an interim analysis, it is useful to express equivalent representations across the scales used in practice and in the plots that follow. Given an observed \(Z_1 = z_1\) with \(n_1\) total observations at the interim analysis, the mappings between scales are:
\begin{align}
\hat{\theta}_1 &= z_1 / \sqrt{n_1} & \text{(standardized effect estimate)} \label{eq:thetahat1}\\
\hat{\delta}_1 &= 2\sigma\hat{\theta}_1 & \text{(unstandardized effect estimate)} \label{eq:deltahat1}\\
\hat{\delta}_1/\delta_1 &= \hat{\theta}_1/\theta_1 & \text{(ratio to planned effect)} \label{eq:ratio1}\\
p_1 &= 1 - \Phi(z_1) & \text{(one-sided } p\text{-value)} \label{eq:pval1}
\end{align}
The information fraction \(t_1 = n_1/N_2\) is fixed by the design; for the group sequential design above, \(t_1 =\) 0.5.
Conditional power given \(Z_1 = z_1\) and \(\theta = \hat{\theta}_1\) follows from equation \eqref{eq:betastar} evaluated at the planned second-stage sample size \(n_2 = N_2 - n_1\):
\begin{equation}
\text{CP}(z_1) = 1 - \Phi\!\left(\frac{b_2 - w_1 z_1}{w_2} - \sqrt{n_2}\,\hat{\theta}_1\right) \label{eq:cp1}
\end{equation}
where \(w_i = \sqrt{n_i/N_2}\), \(i=1,2\), and \(b_2\) is the final efficacy bound from the group sequential design.

The following code illustrates these mappings numerically for the group sequential design derived above. The three x-axis scales in Figure \ref{fig:ssrCP} correspond exactly to the \(Z_1\), \(\hat{\delta}_1/\delta_1\), and \(\text{CP}(z_1)\) columns of this table, evaluated continuously across the interim continuation region \([0.57,\, 2.8)\).

\begin{Shaded}
\begin{Highlighting}[]
\NormalTok{z1\_vals }\OtherTok{\textless{}{-}} \FunctionTok{c}\NormalTok{(}\FloatTok{0.5}\NormalTok{, }\FloatTok{1.0}\NormalTok{, }\FloatTok{1.5}\NormalTok{, }\FloatTok{2.0}\NormalTok{, }\FloatTok{2.5}\NormalTok{)}
\NormalTok{n1      }\OtherTok{\textless{}{-}}\NormalTok{ gsD}\SpecialCharTok{$}\NormalTok{n.I[}\DecValTok{1}\NormalTok{]}
\NormalTok{n2      }\OtherTok{\textless{}{-}}\NormalTok{ gsD}\SpecialCharTok{$}\NormalTok{n.I[}\DecValTok{2}\NormalTok{] }\SpecialCharTok{{-}}\NormalTok{ n1}
\NormalTok{b2      }\OtherTok{\textless{}{-}}\NormalTok{ gsD}\SpecialCharTok{$}\NormalTok{upper}\SpecialCharTok{$}\NormalTok{bound[}\DecValTok{2}\NormalTok{]}
\NormalTok{w1      }\OtherTok{\textless{}{-}} \FunctionTok{sqrt}\NormalTok{(n1 }\SpecialCharTok{/}\NormalTok{ gsD}\SpecialCharTok{$}\NormalTok{n.I[}\DecValTok{2}\NormalTok{])}
\NormalTok{w2      }\OtherTok{\textless{}{-}} \FunctionTok{sqrt}\NormalTok{(n2 }\SpecialCharTok{/}\NormalTok{ gsD}\SpecialCharTok{$}\NormalTok{n.I[}\DecValTok{2}\NormalTok{])}
\FunctionTok{tibble}\NormalTok{(}
  \AttributeTok{Z1         =}\NormalTok{ z1\_vals,}
  \AttributeTok{theta\_hat  =} \FunctionTok{round}\NormalTok{(z1\_vals }\SpecialCharTok{/} \FunctionTok{sqrt}\NormalTok{(n1), }\DecValTok{3}\NormalTok{),}
  \AttributeTok{delta\_hat  =} \FunctionTok{round}\NormalTok{(}\DecValTok{2} \SpecialCharTok{*}\NormalTok{ z1\_vals }\SpecialCharTok{/} \FunctionTok{sqrt}\NormalTok{(n1), }\DecValTok{3}\NormalTok{),  }\CommentTok{\# sigma = 1}
  \AttributeTok{ratio      =} \FunctionTok{round}\NormalTok{(z1\_vals }\SpecialCharTok{/}\NormalTok{ (gsD}\SpecialCharTok{$}\NormalTok{delta }\SpecialCharTok{*} \FunctionTok{sqrt}\NormalTok{(n1)), }\DecValTok{3}\NormalTok{),}
  \AttributeTok{p\_value    =} \FunctionTok{round}\NormalTok{(}\DecValTok{1} \SpecialCharTok{{-}} \FunctionTok{pnorm}\NormalTok{(z1\_vals), }\DecValTok{3}\NormalTok{),}
  \AttributeTok{cond\_pow   =} \FunctionTok{round}\NormalTok{(}\DecValTok{1} \SpecialCharTok{{-}} \FunctionTok{pnorm}\NormalTok{((b2 }\SpecialCharTok{{-}}\NormalTok{ w1 }\SpecialCharTok{*}\NormalTok{ z1\_vals) }\SpecialCharTok{/}\NormalTok{ w2 }\SpecialCharTok{{-}} \FunctionTok{sqrt}\NormalTok{(n2) }\SpecialCharTok{*}\NormalTok{ z1\_vals }\SpecialCharTok{/} \FunctionTok{sqrt}\NormalTok{(n1)), }\DecValTok{3}\NormalTok{)}
\NormalTok{)}
\end{Highlighting}
\end{Shaded}

\begin{verbatim}
## # A tibble: 5 x 6
##      Z1 theta_hat delta_hat ratio p_value cond_pow
##   <dbl>     <dbl>     <dbl> <dbl>   <dbl>    <dbl>
## 1   0.5     0.04      0.081 0.245   0.309    0.036
## 2   1       0.081     0.162 0.49    0.159    0.213
## 3   1.5     0.121     0.243 0.735   0.067    0.581
## 4   2       0.162     0.323 0.98    0.023    0.886
## 5   2.5     0.202     0.404 1.23    0.006    0.986
\end{verbatim}

\section{Code for deriving 2-stage conditional power designs}\label{sec:cpdesigns}

Recall that the above group sequential design was well-powered for the optimistic treatment effect \(\delta = 0.33\), but underpowered for \(\delta = 0.27\). We will attempt to improve power for the lesser treatment effect while maintaining some of the lower sample size benefit of the under-powered group sequential design.
We consider three conditional power variations of the group sequential design shown that are based on boosting conditional power:

\begin{itemize}
\tightlist
\item
  ``Observed treatment effect'' strategy: increase sample size to match conditional power to the originally planned power based on the interim test statistic and observed treatment effect.
\item
  ``Planned treatment effect'' strategy: the same strategy, but assume the originally targeted treatment effect when computing conditional power.
\item
  ``Single adapted N'' strategy: adapt between only two final sample sizes to improve conditional power.
\end{itemize}

The first alternative adapts sample size based on conditional power assuming the observed interim treatment effect and the planned final group sequential design sample size, in this case 154, 306.
In notation, we use the conditional power \(c(z_1, n_2)\) from equation \eqref{eq:czn} where weights are based on \(w_i = \sqrt{n_i / (n_1 + n_2)}\), \(i = 1, 2\).
We also demonstrate \texttt{Power.ssrCP()} to show exact power for \(\delta = 0.27\) is 68.7\%; in the appendix we use \texttt{Power.ssrCP()} to create plots of power and expected sample size and in order to compare designs.

\begin{Shaded}
\begin{Highlighting}[]
\CommentTok{\# Compute adapted total n assuming observed effect size}
\CommentTok{\# and previously derived group sequential design;}
\CommentTok{\# also include enrollment overrun of 75 at interim}
\NormalTok{overrun }\OtherTok{\textless{}{-}} \DecValTok{75}
\NormalTok{n2o }\OtherTok{\textless{}{-}} \FunctionTok{ssrCP}\NormalTok{(}\AttributeTok{z1 =}\NormalTok{ z, }\AttributeTok{x =}\NormalTok{ gsD, }\AttributeTok{cpadj =} \FunctionTok{c}\NormalTok{(}\FloatTok{0.3}\NormalTok{, }\FloatTok{0.8}\NormalTok{), }\AttributeTok{overrun =}\NormalTok{ overrun)}
\CommentTok{\# Print power and expected sample size for delta = .27}
\FunctionTok{Power.ssrCP}\NormalTok{(n2o, }\AttributeTok{theta =} \FloatTok{0.27} \SpecialCharTok{/} \DecValTok{2}\NormalTok{)}
\end{Highlighting}
\end{Shaded}

\begin{verbatim}
##   theta delta     Power       en
## 1 0.135  0.27 0.6868128 330.2952
\end{verbatim}

\begin{Shaded}
\begin{Highlighting}[]
\FunctionTok{plot}\NormalTok{(n2o, }\AttributeTok{ylim =} \FunctionTok{c}\NormalTok{(}\DecValTok{200}\NormalTok{, }\DecValTok{600}\NormalTok{))}
\end{Highlighting}
\end{Shaded}

\begin{figure}

{\centering \includegraphics{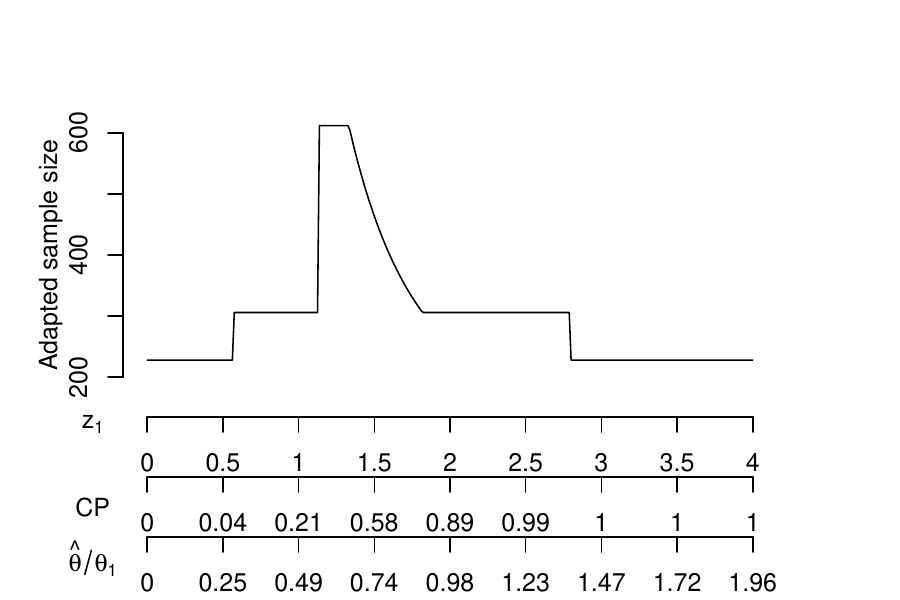} 

}

\caption{Sample size adaptation based on conditional power based on the observed interim treatment effect.}\label{fig:ssrCP}
\end{figure}

Figure \ref{fig:ssrCP} applies the above code to adapt sample size based on the interim test statistic using an extension of the plot command for \texttt{ssrCP} objects. Note the three x-axis scales in the plot: 1) the Z-value (\(Z_1\)) at the interim analysis, 2) the conditional power at the interim analysis for the underlying group sequential design final sample size assuming the interim treatment effect size, and 3) the interim effect size based on the assumption \(Z_1 = \sqrt{n_1}\hat{\theta}_{n_1}\) which will generally hold approximately for asymptotically efficient test statistics. There are 6 intervals on the x-axis to consider on this plot. The two regions on either end represent the interim sample size plus overrun (\texttt{ceiling(gsD\$n.I{[}1{]}\ +\ overrun)} = 228) when the trial stops for futility (\(Z_1 < 0.57\)) or efficacy (\(Z_1 \ge 2.8\)) using the group sequential bounds at the interim analysis:

\begin{Shaded}
\begin{Highlighting}[]
\FunctionTok{round}\NormalTok{(}\FunctionTok{c}\NormalTok{(gsD}\SpecialCharTok{$}\NormalTok{lower}\SpecialCharTok{$}\NormalTok{bound[}\DecValTok{1}\NormalTok{], gsD}\SpecialCharTok{$}\NormalTok{upper}\SpecialCharTok{$}\NormalTok{bound[}\DecValTok{1}\NormalTok{]), }\DecValTok{2}\NormalTok{)}
\end{Highlighting}
\end{Shaded}

\begin{verbatim}
## [1] 0.57 2.80
\end{verbatim}

If we had wished to plot without accounting for the overrun, setting the parameter \texttt{overrun\ =\ 0} would be required.
The intervals just inside these two intervals use the planned final group sequential sample size because the conditional power is outside the interval where sample size is adapted specified by (\texttt{cpadj\ =\ c(0.3,\ 0.8)}) and yet the interim test statistic is in the continuation region for the group sequential design.
In the above code, we set a maximum increase from the planned sample size of no more than 2 times the originally planned sample size (the default \texttt{maxinc\ =\ 2}); this is seen in the region with peak sample size in the plot.\\
With a default of \texttt{z2\ =\ z2NC} we assume a combination test based on the weighted \(Z\)-statistics with weights based on the pre-planned sample sizes as in equation \eqref{eq:wfixed}.
With the CP scale, you can see that the group sequential sample size is adapted up when the conditional power is between 0.3 and 0.8 at the interim analysis.
With the \(\hat{\theta}/\theta_1\) scale, you can see that if the interim treatment effect is approximately as powered for or better (\(\hat{\theta}/\theta_1 \geq 1\)), no sample size adaptation is done.

\begin{figure}

{\centering \includegraphics{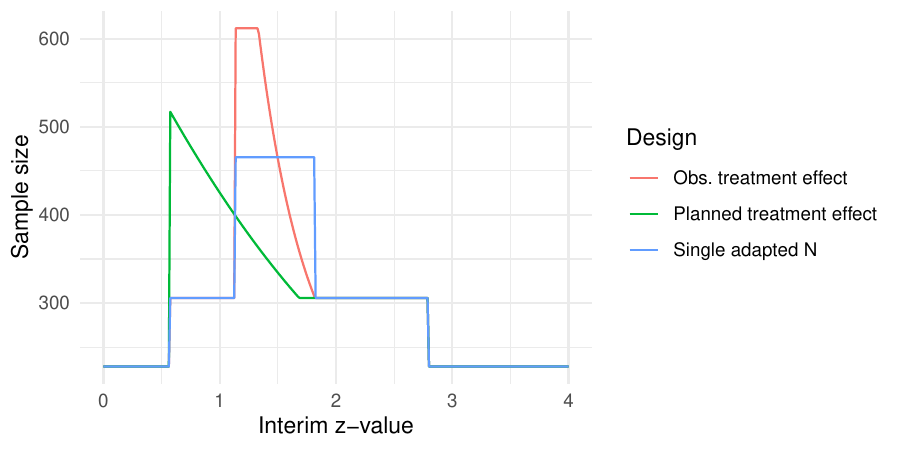} 

}

\caption{Comparison of 3 sample size adaptation strategies.}\label{fig:ssrCP3}
\end{figure}

Next we consider sample size adaptation based on conditional power assuming the originally planned alternate hypothesis treatment effect (\citeproc{ref-LiuChi}{Liu and Chi 2001}).
Note that if we were to use the plot command from above, the CP scale would still be based on the observed interim treatment effect.
The range of conditional power in which sample size will be adapted has been altered to \texttt{cpadj\ =\ c(0.385,\ 0.82)} in order to make the power comparable to the conditional design above as will be seen later in Figure \ref{fig:cppower}. We also set the targeted conditional Type 2 error to 0.177 to match the maximum conditional power for which the sample size is adapted (\(1 - 0.177\)). Note that the maximum increase in sample size is never required for this design (Figure \ref{fig:ssrCP3}).

\begin{Shaded}
\begin{Highlighting}[]
\CommentTok{\# Compute adapted total n assuming H1 effect size}
\NormalTok{n2h1 }\OtherTok{\textless{}{-}} \FunctionTok{ssrCP}\NormalTok{(}
  \AttributeTok{z1 =}\NormalTok{ z, }\AttributeTok{x =}\NormalTok{ gsD, }\AttributeTok{cpadj =} \FunctionTok{c}\NormalTok{(}\FloatTok{0.385}\NormalTok{, }\FloatTok{0.823}\NormalTok{), }\AttributeTok{beta =} \FloatTok{0.177}\NormalTok{,}
  \AttributeTok{overrun =} \DecValTok{75}\NormalTok{, }\AttributeTok{theta =}\NormalTok{ gsD}\SpecialCharTok{$}\NormalTok{theta[}\DecValTok{2}\NormalTok{]}
\NormalTok{)}
\CommentTok{\# Print power and expected sample size for delta = .27}
\FunctionTok{Power.ssrCP}\NormalTok{(n2h1, }\AttributeTok{theta =} \FloatTok{0.27} \SpecialCharTok{/} \DecValTok{2}\NormalTok{)}
\end{Highlighting}
\end{Shaded}

\begin{verbatim}
##   theta delta     Power      en
## 1 0.135  0.27 0.6869699 317.037
\end{verbatim}

For the final conditional power design, we set the targeted Type 2 error at the interim analysis to be small in order to set a single possible sample size to adapt to (Figure \ref{fig:ssrCP3}). The maximum increase is set to 1.522 times the group sequential maximum sample size (\texttt{maxinc\ =\ 1.522}), again to match power for the other conditional power designs (Figure \ref{fig:cppower}).

\begin{Shaded}
\begin{Highlighting}[]
\CommentTok{\# Compute adapted sample size at constant level}
\NormalTok{n2c }\OtherTok{\textless{}{-}} \FunctionTok{ssrCP}\NormalTok{(}\AttributeTok{z1 =}\NormalTok{ z, }\AttributeTok{x =}\NormalTok{ gsD, }\AttributeTok{cpadj =} \FunctionTok{c}\NormalTok{(}\FloatTok{0.3}\NormalTok{, }\FloatTok{0.8}\NormalTok{), }\AttributeTok{overrun =} \DecValTok{75}\NormalTok{,}
             \AttributeTok{beta =} \FloatTok{0.02}\NormalTok{, }\AttributeTok{maxinc =} \FloatTok{1.522}\NormalTok{)}
\CommentTok{\# Print power and expected sample size for delta = .27}
\FunctionTok{Power.ssrCP}\NormalTok{(n2c, }\AttributeTok{theta =} \FloatTok{0.27} \SpecialCharTok{/} \DecValTok{2}\NormalTok{)}
\end{Highlighting}
\end{Shaded}

\begin{verbatim}
##   theta delta     Power       en
## 1 0.135  0.27 0.6868198 327.0911
\end{verbatim}

Figure \ref{fig:ssrCP3} compares the sample size adaptation for the three strategies shown above. In this case, \texttt{ggplot()} is used to plot after combining design characteristics in a data frame as demonstrated below.

\begin{Shaded}
\begin{Highlighting}[]
\CommentTok{\# Add design descriptor to output data frames}
\NormalTok{n2o}\SpecialCharTok{$}\NormalTok{dat}\SpecialCharTok{$}\NormalTok{Design }\OtherTok{\textless{}{-}} \StringTok{"Obs. treatment effect"}
\NormalTok{n2h1}\SpecialCharTok{$}\NormalTok{dat}\SpecialCharTok{$}\NormalTok{Design }\OtherTok{\textless{}{-}} \StringTok{"Planned treatment effect"}
\NormalTok{n2c}\SpecialCharTok{$}\NormalTok{dat}\SpecialCharTok{$}\NormalTok{Design }\OtherTok{\textless{}{-}} \StringTok{"Single adapted N"}
\CommentTok{\# Combine and plot}
\NormalTok{d1 }\OtherTok{\textless{}{-}} \FunctionTok{bind\_rows}\NormalTok{(n2o}\SpecialCharTok{$}\NormalTok{dat, n2h1}\SpecialCharTok{$}\NormalTok{dat, n2c}\SpecialCharTok{$}\NormalTok{dat)}
\FunctionTok{ggplot}\NormalTok{(d1, }\FunctionTok{aes}\NormalTok{(}\AttributeTok{x =}\NormalTok{ z1, }\AttributeTok{y =}\NormalTok{ n2, }\AttributeTok{col =}\NormalTok{ Design)) }\SpecialCharTok{+}
  \FunctionTok{geom\_line}\NormalTok{() }\SpecialCharTok{+}
  \FunctionTok{ylab}\NormalTok{(}\StringTok{"Sample size"}\NormalTok{) }\SpecialCharTok{+}
  \FunctionTok{xlab}\NormalTok{(}\StringTok{"Interim z{-}value"}\NormalTok{) }\SpecialCharTok{+}
  \FunctionTok{theme\_minimal}\NormalTok{()}
\end{Highlighting}
\end{Shaded}

\section{Alternative group sequential designs}\label{sec:altgs}

In addition to the group sequential design used as a basis for the above conditional power designs, we consider two alternatives:

\begin{itemize}
\tightlist
\item
  Group sequential (2): Powered for larger effect size (\(\delta_1 = 0.309\)) to match overall power of ``Observed treatment effect'' conditional power design at \(\delta = 0.27\).
\item
  Group sequential (3): Powered for minimum effect size of interest, \(\delta_1 = 0.27\).
\end{itemize}

\begin{figure}

{\centering \includegraphics{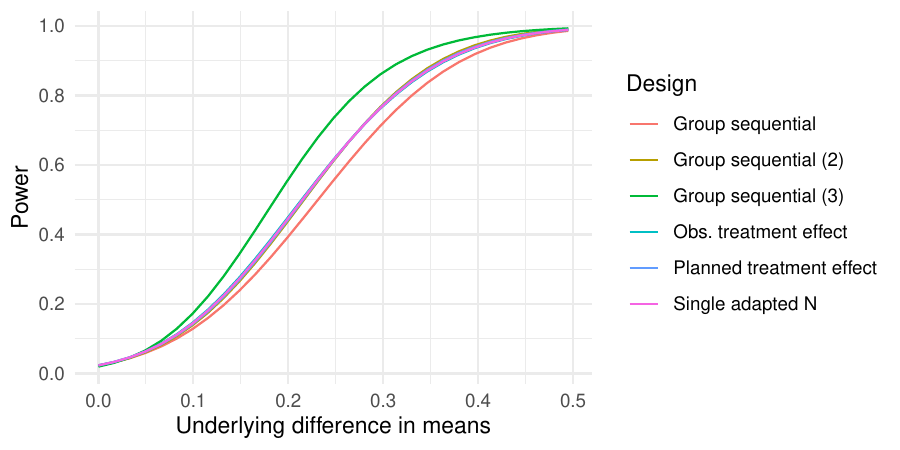} 

}

\caption{Power by underlying difference in treatment means.}\label{fig:cppower}
\end{figure}

\begin{figure}

{\centering \includegraphics{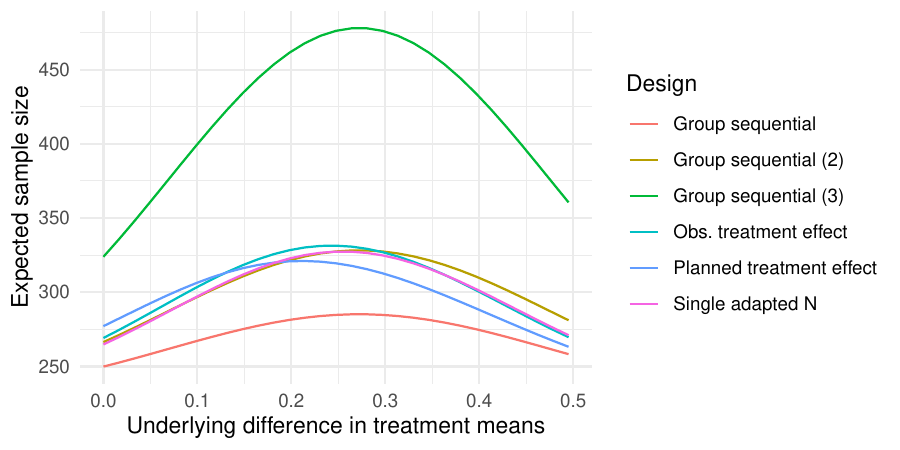} 

}

\caption{Expected sample size for conditional power and group sequential designs.}\label{fig:ploten}
\end{figure}

For both of these alternatives, the interim analysis sample size and bounds are set to be the same as the original group sequential design which means the relative interim timing was decreased from 50\% of observations.
Spending function parameters are set to match the original design \(Z_1\)-cutoff values.
The maximum sample sizes for the 3 group sequential designs are
306,
366, and
570, respectively.
We leave the code for the comparisons in Figures \ref{fig:cppower} and \ref{fig:ploten} to the Appendix since there is some detail involved, but there is not much new in terms of understanding the routines of interest.

The power for these designs and a group sequential design powered at 80\% for \(\delta_1 = 0.27\) (Group sequential (3)) is shown by underlying treatment effect in Figure \ref{fig:cppower}. Like the conditional power designs, the ``Group sequential (2)'' design matches the power of the ``Observed treatment effect'' conditional power adaptive design when the treatment effect is \(\delta = 0.27\), the minimum effect size of interest.
You can immediately see that all of these designs have nearly identical power curves, slightly improved over the original group sequential design. Moving on to Figure \ref{fig:ploten}, we see that while there are subtle differences in the expected sample size curves, none of the designs with similar power curves is uniformly better for expected sample size across the range of treatment effects displayed.
To get a design adequately powered for the lesser treatment effect of \(\delta = 0.27\) requires a fairly substantial increase in expected sample size as displayed using ``Group sequential design (3).''

\section{Summary}\label{sec:summary}

We have summarized the theory behind 2-stage group sequential design and an extension to adapt sample size based on conditional power. Code for deriving and comparing these designs was provided. Plots of designs taking advantage of the ggplot2 functionality are supplied with the gsDesign package and we have shown how to derive and plot conditional power.

For the particular designs shown where the group sequential design is based on an optimistic treatment effect, the conditional power adaptation does not improve overall power to the desired level; this requires a substantially larger sample size as demonstrated using a fully-powered group sequential design. In terms of efficiency of designs with comparable power curves, expected sample sizes are similar and no design is uniformly better than the others across the range of treatment effects examined.
Among designs with similar power and expected sample size, the maximum sample size for the group sequential approach and the ``Single adapted N'' approach seem more attractive than the more traditional conditional power designs.
While we have not generalized these comparisons to other situations, the comparative methods used are easily applied using the provided software.
Certainly the comparisons suggest that naively taking one approach or the other without examining design characteristics carefully may be a mistake.

The fact that design characteristics are approximated well without simulation based on asymptotic theory is convenient in terms of the time that may be required to do a full evaluation of alternative design approaches. The software easily describes the implications of design choices by plotting adaptations based on interim outcomes and also makes it easy to compare power and expected sample sizes for different designs across a range of effect sizes.

While the example we have provided used independent normal observations with a known variance, the theory extends to many endpoint types due to asymptotic theory. For independent, identically distributed observations with an efficient test statistic the theory is straightforward. Thus, the methods demonstrated here can generally be applied for many situations encountered in typical clinical trial design. While survival analysis does not fall into this category, the approach can still be taken with some modification (\citeproc{ref-WassmerCP}{Wassmer 2006}). The functions \texttt{gsDesign()}, \texttt{ssrCP()} and \texttt{condPower()} simplify generation and characterization of 2-stage adaptive sample size using exact calculations and not requiring simulation. The gsDesign package also provides routines that could adapt group sequential designs with multiple interim analyses using either conditional power (\texttt{gsCP()}) or predictive power (\texttt{gsPP()}).
In particular, the methods of Muller and Schafer (\citeproc{ref-MullerSchafer01}{2001}) can be applied by computing the conditional error at an interim analysis using the gsDesign function \texttt{gsCP()} and the remainder of the trial can be designed as an independent trial with Type I error equal to that conditional error with any appropriate design. These designs are not as easily characterized as the 2-stage designs presented here; other R software for such designs is provided by Hack, Brannath, and Brueckner (\citeproc{ref-HackBrannath}{2013}).

\section*{Appendix: Code for Design Comparisons}\label{app:code}
\addcontentsline{toc}{section}{Appendix: Code for Design Comparisons}

The following code computes the power and expected sample size for the conditional power designs and the two alternative group sequential designs used in Figures \ref{fig:cppower} and \ref{fig:ploten}.

The first chunk computes power and expected sample size for the three conditional power designs and for the original group sequential design, and then derives ``Group sequential (2)'' --- a design powered for an intermediate effect size \(\delta_1 = 0.309\) with the same interim sample size and bounds as the original design. The target effect size and spending function parameters are found by fixed-point iteration.

\begin{Shaded}
\begin{Highlighting}[]
\CommentTok{\# Select standardized effect sizes from 0 to 1.5 x H1 effect size}
\NormalTok{thetaplot }\OtherTok{\textless{}{-}}\NormalTok{ (}\DecValTok{0}\SpecialCharTok{:}\DecValTok{30}\NormalTok{) }\SpecialCharTok{/} \DecValTok{20} \SpecialCharTok{*}\NormalTok{ gsD}\SpecialCharTok{$}\NormalTok{theta[}\DecValTok{2}\NormalTok{]}
\CommentTok{\# Compute corresponding parameters on delta (natural parameter) scale}
\NormalTok{deltaplot }\OtherTok{\textless{}{-}}\NormalTok{ thetaplot }\SpecialCharTok{*} \DecValTok{2}
\CommentTok{\# Compute expected sample size and power for conditional power designs}
\NormalTok{en1 }\OtherTok{\textless{}{-}} \FunctionTok{Power.ssrCP}\NormalTok{(n2o, }\AttributeTok{theta =}\NormalTok{ thetaplot)}
\NormalTok{en2 }\OtherTok{\textless{}{-}} \FunctionTok{Power.ssrCP}\NormalTok{(n2h1, }\AttributeTok{theta =}\NormalTok{ thetaplot)}
\NormalTok{en3 }\OtherTok{\textless{}{-}} \FunctionTok{Power.ssrCP}\NormalTok{(n2c, }\AttributeTok{theta =}\NormalTok{ thetaplot)}
\CommentTok{\# Add additional effect sizes to original group sequential design}
\NormalTok{gsDx }\OtherTok{\textless{}{-}} \FunctionTok{gsProbability}\NormalTok{(}\AttributeTok{d =}\NormalTok{ gsD, }\AttributeTok{theta =}\NormalTok{ thetaplot)}
\NormalTok{en4 }\OtherTok{\textless{}{-}} \FunctionTok{data.frame}\NormalTok{(}
  \AttributeTok{theta =}\NormalTok{ thetaplot, }\AttributeTok{en =}\NormalTok{ gsDx}\SpecialCharTok{$}\NormalTok{en,}
  \AttributeTok{Power =} \FunctionTok{as.vector}\NormalTok{(gsDx}\SpecialCharTok{$}\NormalTok{upper}\SpecialCharTok{$}\NormalTok{prob[}\DecValTok{1}\NormalTok{, ] }\SpecialCharTok{+}\NormalTok{ gsDx}\SpecialCharTok{$}\NormalTok{upper}\SpecialCharTok{$}\NormalTok{prob[}\DecValTok{2}\NormalTok{, ]),}
  \AttributeTok{delta =}\NormalTok{ deltaplot, }\AttributeTok{Design =} \StringTok{"Group sequential"}
\NormalTok{)}

\CommentTok{\# Compute group sequential design powered for a larger effect size}
\NormalTok{deltanew }\OtherTok{\textless{}{-}} \FloatTok{0.3095}
\NormalTok{n.fix }\OtherTok{\textless{}{-}} \FunctionTok{nNormal}\NormalTok{(}\AttributeTok{delta1 =}\NormalTok{ deltanew, }\AttributeTok{sd =} \DecValTok{1}\NormalTok{, }\AttributeTok{beta =} \FloatTok{0.2}\NormalTok{, }\AttributeTok{alpha =} \FloatTok{0.025}\NormalTok{)}
\NormalTok{overrun }\OtherTok{\textless{}{-}} \DecValTok{75}
\NormalTok{timing }\OtherTok{\textless{}{-}} \FloatTok{0.4255}
\NormalTok{gsDa }\OtherTok{\textless{}{-}} \FunctionTok{gsProbability}\NormalTok{(}\AttributeTok{d =}\NormalTok{ gsD, }\AttributeTok{theta =}\NormalTok{ deltanew }\SpecialCharTok{/} \DecValTok{2}\NormalTok{)}
\NormalTok{sflprob }\OtherTok{\textless{}{-}}\NormalTok{ gsDa}\SpecialCharTok{$}\NormalTok{lower}\SpecialCharTok{$}\NormalTok{prob[}\DecValTok{1}\NormalTok{, }\DecValTok{1}\NormalTok{]}

\CommentTok{\# Fixed point iteration to match interim bounds and timing}
\NormalTok{n }\OtherTok{\textless{}{-}}\NormalTok{ (deltanew }\SpecialCharTok{/} \FloatTok{0.27}\NormalTok{)}\SpecialCharTok{\^{}}\DecValTok{2} \SpecialCharTok{*}\NormalTok{ gsD}\SpecialCharTok{$}\NormalTok{n.I[}\DecValTok{2}\NormalTok{]}
\NormalTok{sfupar }\OtherTok{\textless{}{-}}\NormalTok{ gsD}\SpecialCharTok{$}\NormalTok{upper}\SpecialCharTok{$}\NormalTok{param}
\NormalTok{sflpar }\OtherTok{\textless{}{-}}\NormalTok{ gsD}\SpecialCharTok{$}\NormalTok{lower}\SpecialCharTok{$}\NormalTok{param}
\NormalTok{nI }\OtherTok{\textless{}{-}}\NormalTok{ n}
\ControlFlowTok{while}\NormalTok{ (}\FunctionTok{abs}\NormalTok{(nI }\SpecialCharTok{{-}}\NormalTok{ gsD}\SpecialCharTok{$}\NormalTok{n.I[}\DecValTok{1}\NormalTok{]) }\SpecialCharTok{\textgreater{}} \FloatTok{0.01}\NormalTok{) \{}
\NormalTok{  timing }\OtherTok{\textless{}{-}}\NormalTok{ gsD}\SpecialCharTok{$}\NormalTok{n.I[}\DecValTok{1}\NormalTok{] }\SpecialCharTok{/}\NormalTok{ n}
\NormalTok{  gsD2 }\OtherTok{\textless{}{-}} \FunctionTok{gsDesign}\NormalTok{(}
    \AttributeTok{k =} \DecValTok{2}\NormalTok{, }\AttributeTok{timing =}\NormalTok{ timing, }\AttributeTok{beta =} \FloatTok{0.2}\NormalTok{, }\AttributeTok{alpha =} \FloatTok{0.025}\NormalTok{, }\AttributeTok{delta1 =}\NormalTok{ deltanew,}
    \AttributeTok{n.fix =}\NormalTok{ n.fix, }\AttributeTok{overrun =}\NormalTok{ overrun,}
    \AttributeTok{sfu =}\NormalTok{ sfPower, }\AttributeTok{sfl =}\NormalTok{ sfPower, }\AttributeTok{sfupar =}\NormalTok{ sfupar, }\AttributeTok{sflpar =}\NormalTok{ sflpar}
\NormalTok{  )}
\NormalTok{  sfupar }\OtherTok{\textless{}{-}}\NormalTok{ (}\FunctionTok{log}\NormalTok{(gsD}\SpecialCharTok{$}\NormalTok{upper}\SpecialCharTok{$}\NormalTok{prob[}\DecValTok{1}\NormalTok{, }\DecValTok{1}\NormalTok{]) }\SpecialCharTok{{-}} \FunctionTok{log}\NormalTok{(}\FloatTok{0.025}\NormalTok{)) }\SpecialCharTok{/} \FunctionTok{log}\NormalTok{(timing)}
\NormalTok{  gsDa }\OtherTok{\textless{}{-}} \FunctionTok{gsProbability}\NormalTok{(}\AttributeTok{d =}\NormalTok{ gsD2, }\AttributeTok{theta =}\NormalTok{ deltanew }\SpecialCharTok{/} \DecValTok{2}\NormalTok{)}
\NormalTok{  sflpar }\OtherTok{\textless{}{-}}\NormalTok{ (}\FunctionTok{log}\NormalTok{(sflprob) }\SpecialCharTok{{-}} \FunctionTok{log}\NormalTok{(}\FloatTok{0.2}\NormalTok{)) }\SpecialCharTok{/} \FunctionTok{log}\NormalTok{(timing)}
\NormalTok{  n }\OtherTok{\textless{}{-}}\NormalTok{ gsD2}\SpecialCharTok{$}\NormalTok{n.I[}\DecValTok{2}\NormalTok{]}
\NormalTok{  nI }\OtherTok{\textless{}{-}}\NormalTok{ gsD2}\SpecialCharTok{$}\NormalTok{n.I[}\DecValTok{1}\NormalTok{]}
\NormalTok{\}}
\NormalTok{gsDx2 }\OtherTok{\textless{}{-}} \FunctionTok{gsProbability}\NormalTok{(}\AttributeTok{d =}\NormalTok{ gsD2, }\AttributeTok{theta =}\NormalTok{ thetaplot)}
\NormalTok{en5 }\OtherTok{\textless{}{-}} \FunctionTok{data.frame}\NormalTok{(}
  \AttributeTok{theta =}\NormalTok{ thetaplot, }\AttributeTok{en =}\NormalTok{ gsDx2}\SpecialCharTok{$}\NormalTok{en,}
  \AttributeTok{Power =} \FunctionTok{as.vector}\NormalTok{(gsDx2}\SpecialCharTok{$}\NormalTok{upper}\SpecialCharTok{$}\NormalTok{prob[}\DecValTok{1}\NormalTok{, ] }\SpecialCharTok{+}\NormalTok{ gsDx2}\SpecialCharTok{$}\NormalTok{upper}\SpecialCharTok{$}\NormalTok{prob[}\DecValTok{2}\NormalTok{, ]),}
  \AttributeTok{delta =}\NormalTok{ deltaplot, }\AttributeTok{Design =} \StringTok{"Group sequential (2)"}
\NormalTok{)}
\end{Highlighting}
\end{Shaded}

The second chunk derives ``Group sequential (3)'', which is fully powered at 80\% for the minimum effect size of interest \(\delta = 0.27\), again matching the original design's interim sample size and bounds.

\begin{Shaded}
\begin{Highlighting}[]
\CommentTok{\# Computations for design fully powered for delta = .27}
\NormalTok{nnor3 }\OtherTok{\textless{}{-}} \FunctionTok{nNormal}\NormalTok{(}\AttributeTok{delta1 =} \FloatTok{0.27}\NormalTok{, }\AttributeTok{sd =} \DecValTok{1}\NormalTok{, }\AttributeTok{beta =} \FloatTok{0.2}\NormalTok{, }\AttributeTok{alpha =} \FloatTok{0.025}\NormalTok{)}
\NormalTok{thetanew }\OtherTok{\textless{}{-}} \FunctionTok{gsDesign}\NormalTok{(}\AttributeTok{beta =} \FloatTok{0.2}\NormalTok{, }\AttributeTok{n.fix =}\NormalTok{ nnor3)}\SpecialCharTok{$}\NormalTok{delta}
\NormalTok{gsDa }\OtherTok{\textless{}{-}} \FunctionTok{gsProbability}\NormalTok{(}\AttributeTok{d =}\NormalTok{ gsD, }\AttributeTok{theta =}\NormalTok{ thetanew)}
\NormalTok{timing }\OtherTok{\textless{}{-}} \FloatTok{0.269}
\NormalTok{sfupar }\OtherTok{\textless{}{-}}\NormalTok{ (}\FunctionTok{log}\NormalTok{(gsD}\SpecialCharTok{$}\NormalTok{upper}\SpecialCharTok{$}\NormalTok{prob[}\DecValTok{1}\NormalTok{, }\DecValTok{1}\NormalTok{]) }\SpecialCharTok{{-}} \FunctionTok{log}\NormalTok{(}\FloatTok{0.025}\NormalTok{)) }\SpecialCharTok{/} \FunctionTok{log}\NormalTok{(timing)}
\NormalTok{sflpar }\OtherTok{\textless{}{-}}\NormalTok{ (}\FunctionTok{log}\NormalTok{(gsDa}\SpecialCharTok{$}\NormalTok{lower}\SpecialCharTok{$}\NormalTok{prob[}\DecValTok{1}\NormalTok{, }\DecValTok{1}\NormalTok{]) }\SpecialCharTok{{-}} \FunctionTok{log}\NormalTok{(}\FloatTok{0.2}\NormalTok{)) }\SpecialCharTok{/} \FunctionTok{log}\NormalTok{(timing)}

\NormalTok{gsD3 }\OtherTok{\textless{}{-}} \FunctionTok{gsDesign}\NormalTok{(}
  \AttributeTok{k =} \DecValTok{2}\NormalTok{, }\AttributeTok{timing =}\NormalTok{ timing, }\AttributeTok{beta =} \FloatTok{0.2}\NormalTok{, }\AttributeTok{alpha =} \FloatTok{0.025}\NormalTok{, }\AttributeTok{delta1 =} \FloatTok{0.27}\NormalTok{,}
  \AttributeTok{n.fix =} \FunctionTok{nNormal}\NormalTok{(}\AttributeTok{delta1 =} \FloatTok{0.27}\NormalTok{, }\AttributeTok{sd =} \DecValTok{1}\NormalTok{, }\AttributeTok{beta =} \FloatTok{0.2}\NormalTok{, }\AttributeTok{alpha =} \FloatTok{0.025}\NormalTok{),}
  \AttributeTok{overrun =} \DecValTok{75}\NormalTok{, }\AttributeTok{sfu =}\NormalTok{ sfPower, }\AttributeTok{sfl =}\NormalTok{ sfPower, }\AttributeTok{sfupar =}\NormalTok{ sfupar, }\AttributeTok{sflpar =}\NormalTok{ sflpar}
\NormalTok{)}
\NormalTok{gsDx3 }\OtherTok{\textless{}{-}} \FunctionTok{gsProbability}\NormalTok{(}\AttributeTok{d =}\NormalTok{ gsD3, }\AttributeTok{theta =}\NormalTok{ thetaplot)}
\NormalTok{en6 }\OtherTok{\textless{}{-}} \FunctionTok{data.frame}\NormalTok{(}
  \AttributeTok{theta =}\NormalTok{ thetaplot, }\AttributeTok{en =}\NormalTok{ gsDx3}\SpecialCharTok{$}\NormalTok{en,}
  \AttributeTok{Power =} \FunctionTok{as.vector}\NormalTok{(gsDx3}\SpecialCharTok{$}\NormalTok{upper}\SpecialCharTok{$}\NormalTok{prob[}\DecValTok{1}\NormalTok{, ] }\SpecialCharTok{+}\NormalTok{ gsDx3}\SpecialCharTok{$}\NormalTok{upper}\SpecialCharTok{$}\NormalTok{prob[}\DecValTok{2}\NormalTok{, ]),}
  \AttributeTok{delta =}\NormalTok{ deltaplot, }\AttributeTok{Design =} \StringTok{"Group sequential (3)"}
\NormalTok{)}
\NormalTok{en1}\SpecialCharTok{$}\NormalTok{Design }\OtherTok{\textless{}{-}} \StringTok{"Obs. treatment effect"}
\NormalTok{en2}\SpecialCharTok{$}\NormalTok{Design }\OtherTok{\textless{}{-}} \StringTok{"Planned treatment effect"}
\NormalTok{en3}\SpecialCharTok{$}\NormalTok{Design }\OtherTok{\textless{}{-}} \StringTok{"Single adapted N"}
\end{Highlighting}
\end{Shaded}

\section*{References}\label{references}
\addcontentsline{toc}{section}{References}

\protect\phantomsection\label{refs}

\end{document}